\RequirePackage{fix-cm}
\documentclass[twocolumn]{svjour3}          
\smartqed  
\usepackage{graphicx}

\usepackage{times}
\usepackage{subfigure}
\usepackage[hyphens]{url}

\usepackage{enumitem}

\usepackage[
bookmarksnumbered = true,
bookmarksopen = true,
bookmarksopenlevel=3,
pdftitle={Performance Impact of Data Layout on the GPU-accelerated IDW Interpolation},
pdfauthor={Gang Mei},
pdfstartview = FitH,
colorlinks=true,
linkcolor=blue,
breaklinks=true,
urlcolor=black,
citecolor=blue
]{hyperref}

\begin{document}

\title{Performance Impact of Data Layout on the GPU-accelerated \\IDW Interpolation
}


\author{Gang Mei         \and
        Hong Tian   
}


\institute{G. Mei\at
              Institute of Earth and Environmental Science, University of Freiburg, Albertstr.23B, D-79104, Freiburg im Breisgau, Germany\\
              Tel.: +49 761 203 6477\\
              Fax: +49 761 203 6496\\
              \email{gangmeiphd@gmail.com} \\
           \and
           H. Tian \at
              Faculty of Engineering, China University of Geosciences (Wuhan), \\ 
              No.388 Lumo Road, 430074, Wuhan, China\\
              \email{htian2011@hotmail.com}
}

\date{Received: date / Accepted: date}

\maketitle

\begin{abstract}
This paper focuses on evaluating the performance impact of different data 
layouts on the GPU-accelerated IDW interpolation. First, we redesign and improve our 
previous GPU implementation that was performed by exploiting the feature 
CUDA Dynamic Parallel (CDP). And then, we 
implement three versions of GPU implementations, i.e., the na\"{\i}ve 
version, the tiled version, and the improved CDP version, based on five 
layouts including the \textit{Structure of Arrays} (SoA), the \textit{Array of Sturcutes} (AoS), the \textit{Array of aligned Sturcutes} (AoaS), the \textit{Structure of Arrays of aligned Structures} (SoAoS), and the 
\textit{Hybrid} layout. Experimental results show that: the layouts AoS and AoaS achieve 
better performance than the layout SoA for both the na\"{\i}ve version and 
tiled version, while the layout SoA is the best choice for the improved CDP 
version. We also observe that: for the two combined data layouts (the SoAoS 
and the Hybrid), there are no notable performance gains when compared to 
other three basic layouts. We recommend that: in practical applications, the 
layout AoaS is the best choice since the tiled version is the fastest one 
among the three versions of GPU implementations, especially on single 
precision. 
\keywords{GPU \and Data Layout \and IDW Interpolation \and  CUDA Dynamic Parallelism}
 
\end{abstract}

\section{Introduction}
\label{sec:introduction}

Data layout is the form in which data should be organized and accessed in 
memory when operating on multi-valued data such as sets of 3D points. The 
selecting of appropriate data layout is a crucial issue in the development 
of GPU-accelerated applications. The efficiency performance of the same GPU 
application may drastically differs due to the use of different types of 
data layout; see the example of sorting structures demonstrated with 
\textit{Thrust} \cite{bell2011}. 

Typically, there are two major choices of the data layout: the 
\textit{Array-of-Structures} (AoS) and the \textit{Structure-of-Arrays} (SoA) \cite{farber2011}; see Figure \ref{fig:data:layout}. Organizing data in AoS layout 
leads to coalescing issues as the data are interleaved. In contrast, the 
organizing of data according to the SoA layout can generally make full use 
of the memory bandwidth due to no data interleaving. In addition, global 
memory accesses based upon the SoA layout are always coalesced. The above 
two layouts are probably the most basic and simplest memory access patterns.
More complex data layouts such as AoSoA \cite{abel1999} and SoAoS \cite{siegel2009} can be formed 
by combining the basic layouts AoS and SoA.

As noted above, the memory access patterns are critical 
for the performance of GPU-accelerated applications. However, it is not 
always obvious which data layout will achieve better performance for a 
specific application. For example, in order to evaluate the performance of 
the SoA and AoS layouts, Govender, et al. \cite{govender2013}  ran a 
simulation of 2 million particles using their discrete element simulation 
framework, and found that AoS is three times slower than SoA, while an 
opposite argument was presented in \cite{giles2013}. In the library 
framework OP2, Giles, et al. \cite{giles2013} preferred to use the AoS 
layout to store mesh data for better memory accesses performance. In 
practice, a common solution is to implement a specific application using 
above two layouts separately and then compare the performance. 

The Inverse Distance Weighting (IDW) interpolation algorithm, which was 
originally proposed by Shepard \cite{shepard1968}, is one of the most commonly used 
spatial interpolation methods in Geosciences. Typically, the implementation 
of spatial interpolation within the conventional sequential programming 
patterns is computationally expensive for a large number of data sets. In 
order to improve the computational efficiency, some efforts have been 
carried out to develop efficient implementations of the IDW interpolation in 
various massively parallel computing environments on multi-core CPUs 
\cite{armstrong1997,guan2010,huang2011} and/or GPUs platforms \cite{hanzer2012,hennebohl2011,huraj2010,xia2011}.

In our previous work \cite{mei2014}, we presented two GPU implementations of the 
standard IDW interpolation algorithm, the tiled version that took advantage of shared 
memory and the CDP version that was implemented by exploiting CUDA Dynamic 
Parallelism (CDP). We found that the tilted version achieved the highest speedups 
over the CPU version. However, the CDP version is 4.8x $\sim $ 6.0x slower 
than the na\"{\i}ve GPU version. Those experimental tests were 
performed only on single precision.

In this paper, we focus on evaluating the performance impact of different 
data layouts when implementing the IDW interpolation on the GPU. We first 
redesign the CDP version to avoid the use of the atomic operation 
\texttt{atomicAdd()}, and then test three GPU implementations, i.e., the na\"{\i}ve 
GPU version presented in \cite{huraj2010}, the tiled version described in \cite{mei2014}, and 
the improved CDP version introduced in this paper, on both single precision 
and double precision. In our previous work \cite{mei2014}, the above three GPU 
implementations are developed according to the data layout SoA. In order to 
evaluate the impact of other data layouts such as AoS, we also implement 
these GPU versions based upon the AoS layout and other combined data layouts 
such as SoAoS \cite{siegel2009}, and then test their performance on single and double 
precision.

In summary, we make the following contributions in this paper:

\begin{enumerate}[label=(\arabic*)]
\item Redesign the CDP version that is originally presented in \cite{mei2014} to improve its efficiency.

\item Implement several groups of those three GPU versions based upon several different data layouts
on both single and double precision. 

\item Evaluate the performance of sets of GPU implementations that are developed 
according to several different layouts on both single and double precision.
\end{enumerate}

This paper is organized as follows. Section \ref{sec:background} gives a brief 
introduction to the IDW interpolation and two basic 
data layouts, the SoA and AoS. Section \ref{sec:implement} concentrates on the 
GPU implementations that are performed  by using five different 
data layouts. Section \ref{sec:result:and:discuss} presents some experimental tests that are performed 
on both single and double precision, and discusses the experimental results. Finally, 
Section \ref{sec:conclusion} draws some conclusions.

\section{Background}
\label{sec:background}

\subsection{IDW Interpolation}
\label{sec:back:idw}

The IDW algorithm is one of the most commonly used spatial interpolation 
methods in Geosciences, which calculates the interpolated values of unknown 
points (prediction points) by weighting average of the values of known 
points (data points). The name given to this type of methods was motivated 
by the weighted average applied since it resorts to the inverse of the 
distance to each known point when calculating the weights. The difference 
between different forms of IDW interpolation is that they calculate the 
weights variously. 

A general form of predicting an interpolated value $Z$ at a given point $x$ based 
on samples $Z_{i}=Z(x_{i})$ for $i$ = 1, 2, {\ldots}, $n$ using IDW is an 
interpolating function: 

\begin{equation}
\label{eq1}
Z(x)=\sum\limits_{i=1}^n {\frac{\omega _i (x)z_i }{\sum\limits_{j=1}^n 
{\omega _j (x)} }} ,
\quad
\omega _i (x)=\frac{1}{d(x,x_i )^p}
\end{equation}

The above equation is a simple IDW weighting function, as defined by 
Shepard \cite{shepard1968}, where $x$ denotes a predication location, $x_{i }$ is a data 
point, $d$ is the distance from the known data point $x_{i}$ to the unknown 
prediction point $x$, $n$ is the total number of data points used in 
interpolating, and $p$ is an arbitrary positive real number called the power 
parameter (typically, $p = 2$).

\subsection{Data Layout}
\label{sec:back:layout}

\begin{figure}[htb]
    \centering
  \subfigure[][AoS]{
    \label{fig:data:layout:aos}       
    \includegraphics[scale = 0.6]{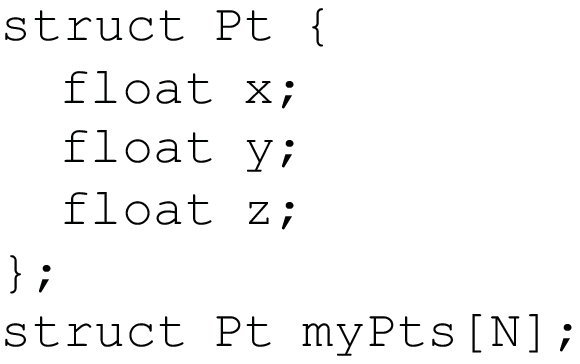}
    }
    \hspace{1em}
  \subfigure[SoA]{
    \label{fig:data:layout:soa}       
    \includegraphics[scale = 0.6]{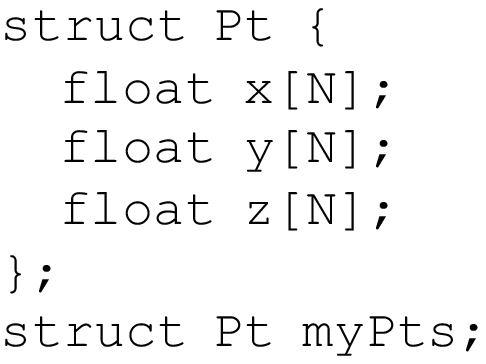}
    }
    \caption{Data layouts: Array-of-Structures (AoS) and Structure-of-Arrays (SoA)}
    \label{fig:data:layout}       
\end{figure}

In GPU computing, an optimal pattern of accessing data can significantly 
improve the overall efficiency performance by minimizing the number of 
memory transactions on the off-chip global memory. Thus, one of the key 
design issues for generating efficient GPU code is the selecting of proper 
data layouts when operating on multi-valued data such as sets of points or 
pixels. In general, there are two major choices of the data layout: the 
\textit{Array-of-Structures} (AoS) and the \textit{Structure-of-Arrays} (SoA); see Figure \ref{fig:data:layout}. 

Organizing data in AoS layout leads to coalescing issues as the data are 
interleaved. Multi-dimensional and multi-valued data containers lead to 
strided memory accesses in the one dimensional address space, and cause 
exactly this problem. For example, performing an operation on a set of 3D 
points illustrated in Figure \ref{fig:data:layout:aos} that only requires the variable $x$ will 
result in about a 66 percent loss of bandwidth and waste of L2 cache memory.

In contrast, the organizing of data according to the SoA layout can 
typically make full use of the memory bandwidth since there is no data 
interleaving; see Figure \ref{fig:data:layout:soa}. Furthermore, global memory accesses are always coalesced when 
using this type of data layout; and usually higher global memory performance 
can be achieved. 

The SoA data layout is beneficial in many cases. Farber \cite{farber2011} suggested 
that from a GPU performance perspective, it is preferable to use the SoA 
layout. This argument was demonstrated by the example of sorting SoA and AoS 
structures with \textit{Thrust}; it was reported that a 5-times speedup can be 
achieved by using a SoA data structure over a AoS data structure \cite{bell2011}.

Similarly, in order to gauge the effective performance of the two 
representations, i.e., the SoA and AoS layouts, on the GPU, 
Govender, et al. \cite{govender2013} ran a simulation of 2 million particles 
using their discrete element simulation framework BLAZE-DEM, and found that 
AoS is three times slower than SoA.

However, an opposite argument was presented in \cite{giles2013}. In the library 
framework for the solution of unstructured mesh applications OP2, Giles, et al. \cite{giles2013} and Mudalige, et al. 
\cite{mudalige2013} preferred to use the AoS layout to store mesh data for better memory 
accesses performance. 

The above mentioned applications indicate that memory access patterns (e.g., AoS 
and SoA) are critical for performance, especially on parallel architectures 
such as GPUs. However, it is not always obvious which data layout will 
achieve better performance in a particular application. The selection of a proper data layout for a specific application depends on its underlying algorithm. In general, the 
usual language syntax and standard container types lead naturally to the AoS 
layout while SIMD units much prefer the SoA format \cite{strzodka2012}. 

In order to improve the efficiency of accessing memories on the GPU, many 
studies have been performed to transform different types of layouts to 
others, e.g., from AoS to SoA, or vice versa  \cite{mistry2011,strzodka2011,strzodka2012,sung2012}. Furthermore, the 
major choices of AoS and SoA can be further refined to form hybrid formats, 
e.g., arrays of structures of arrays \cite{abel1999} or structures of arrays of 
structures \cite{siegel2009}.

In this work, we will evaluate the performance impact of the above two basic data layouts and other layouts that are derived from the above two layouts. A group of GPU implementations of those three versions will be developed particularly by using one type of data layout, and then compared to other groups of implementations. 

\section{GPU Implementations}
\label{sec:implement}

\subsection{The SoA Group of Implementations}
\label{sec:implement:soa}

In our previous work \cite{mei2014}, we have introduced three GPU implementations 
of the standard IDW interpolations, i.e., the na\"{\i}ve version, the tiled 
version, and the CDP version. These GPU implementations are completely 
developed according to the data layout SoA. In \cite{mei2014}, we also tested these 
three versions using several sets of data on single precision. 

In this section, we first describe an improved version of the CDP 
implementation. The CDP version presented in \cite{mei2014}, which is referred to 
as the original CDP version in this section, has two levels of nested 
parallelism: (1) level 1: for all prediction points, the interpolated values 
can be calculated in parallel; (2) lever 2: for each prediction point, the 
distances to all data points can be calculated in parallel. The parent 
kernel is responsible for performing the first level of parallelism, while 
the child kernel takes responsibility for realizing the second level of 
parallelism.

In the original CDP version, each child kernel is responsible for 
calculating the distances from all data points to a predication point. More 
specifically, first each thread within a child grid is invoked for 
calculating: (1) the distance from one data point to a predication point, 
(2) the corresponding weight, and (3) the weighted value (see Equation (\ref{eq1})); 
and then the weights and weighted values calculated within the same thread 
block will be locally accumulated using the parallel reduction \cite{harris2007}; 
finally all weights and weighted values that have been obtained within 
different blocks will be accumulated using the atomic operation \texttt{atomicAdd()}.

The atomic operations such as \texttt{atomicAdd()} cannot be performed on double 
precision. In order to enable the CDP version to be executed on double 
precision, we redesign and improve this GPU implementation to avoid the use of 
\texttt{atomicAdd()}. The basic idea behind this improvement is as follows.

In the improved CDP version, we no longer allocate $n$ threads within a child 
grid (where $n$ is the number of data points), but  only allocate one thread 
block with 1024 threads. Within this single thread block, each thread is 
responsible for calculating the distances of several data points rather than 
only one data point to a predication point. For example, assuming there are 
3000 data points, for each predication point, it is needed to calculate all 
the distances from the predication point to those 3000 data points. Each 
thread will take responsibilities for calculating three (i.e., $(3000 + 1024 - 1) 
/ 1024$) distances. These three distances and corresponding weights will be 
locally accumulated within each thread; and when all threads within the only 
one block finish calculating all distances, the accumulation of all weights 
and weighted values will be achieved by performing a parallel reduction 
\cite{harris2007} within the thread block. Thus, in this situation, the operation 
\texttt{atomicAdd()} is not needed for accumulating all weights and weighted values 
that have been calculated within different blocks of threads. 

We test the performance of the improved CDP version using five sets of data. In each set of test data, the numbers of data points and predication points are to be identical. We create five groups of sizes, i.e., 10K, 50K, 100K, 500K, and 1000K (1K=1024). And five tests are performed by setting the numbers of both the data points and prediction points as the above listed five groups of sizes. 

\begin{figure}[htb]
    \centering
  \subfigure[Execution time of the new and old versions]{
    \label{fig:new:cdp:a}       
    \includegraphics[scale = 0.35]{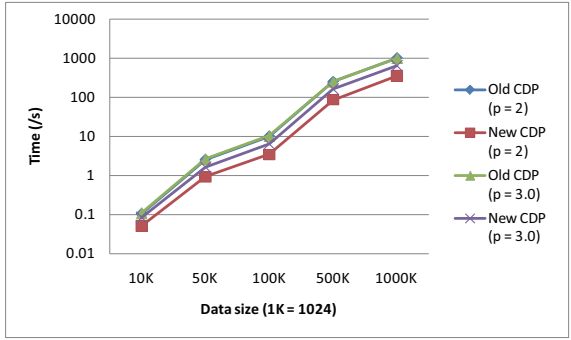}
    }
    \hspace{1em}
  \subfigure[Speedups of the new version over the old version]{
    \label{fig:new:cdp:b}       
    \includegraphics[scale = 0.35]{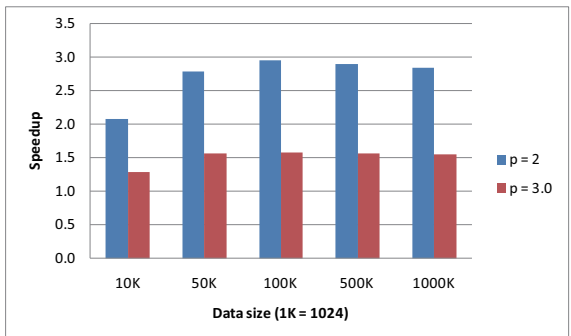}
    }
    \caption{Performance comparison of the original (old) and  the improved (new) CDP versions}
    \label{fig:new:cdp}       
\end{figure}

The performance of the original and the improved CDP versions is illustrated in Figure \ref{fig:new:cdp}. These experimental tests show that the improved CDP version achieves the speedups of 2.9x and 1.5x over the original CDP version when the power parameter $p$ is set to 2 and 3.0, respectively. Noticeably, for the original CDP version, the performance in the two cases where the power parameter $p$ is set as 2 and 3.0 is almost the same; thus, in the Figure \ref{fig:new:cdp:a}, the two lines representing the execution time of the old version are almost overlapped.

In this paper, the na\"{\i}ve version presented in \cite{huraj2010}, the tiled version 
developed in \cite{mei2014}, and the improved CDP version described above are 
accepted to be implemented according to different data layouts for benchmark 
tests on both single precision and double precision. 

\subsection{The AoS Group of Implementations}
\label{sec:implement:aos}

The implementations of the three GPU versions according to the layout AoS is 
quite straightforward, which can be realized by simply modifying the SoA 
group of the three GPU implementations. First, two arrays of structures 
that are used to store the coordinates of all data points and predications 
points are allocated; and then the references to the points' coordinates in 
the SoA version of the GPU implementations are replaced by using the arrays 
that are represented in the AoS format. 

Note that, in this group of implementations, the structure for 
representing 3D points are misaligned; in other words, the structure is 
not forced to be aligned using the specifier \texttt{{\_}{\_}align{\_}{\_}()}; see 
Figure \ref{fig:data:layout:aos}.

\subsection{The AoaS Group of Implementations}
\label{sec:implement:aoas}

In the AoS group of GPU implementations described above, the data structure for representing 3D 
points is not forced to be aligned. Operations using the misaligned 
structure may requires much more memory transactions when accessing global 
memory, and thus decreases the overall efficiency performance \cite{cuda2013}.

\begin{figure}[htb]
    \centering
  \subfigure[Single precision]{
    \label{fig:layout:aoas:a}       
    \includegraphics[scale = 0.6]{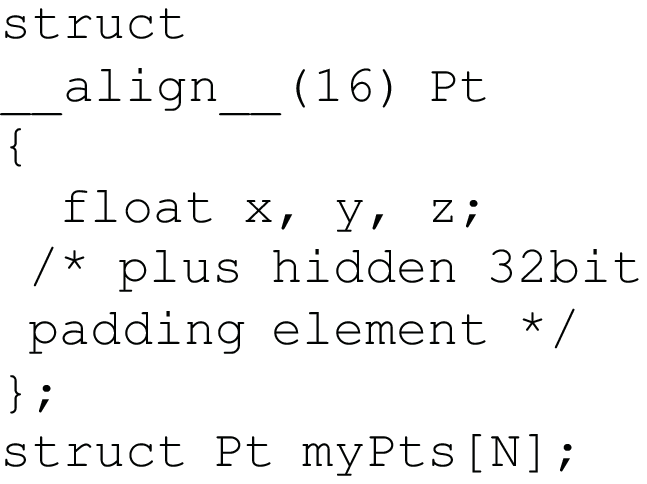}
    }
  \subfigure[Double precision]{
    \label{fig:layout:aoas:b}       
    \includegraphics[scale = 0.6]{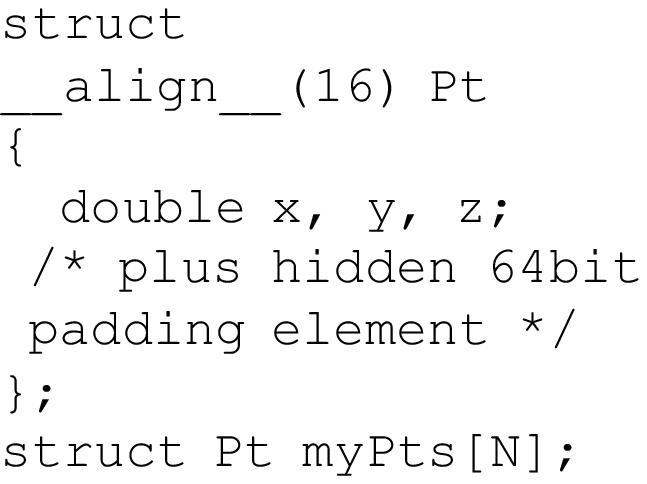}
    }
    \caption{Data layout: Array of aligned Structures (AoaS)}
    \label{fig:layout:aoas}       
\end{figure}

In order to benefit from the aligned memory accesses, we simply add the 
specifier \texttt{{\_}{\_}align{\_}{\_}} into the data structures; see Figure \ref{fig:layout:aoas}. 
Noticeably, on single precision, a hidden 32 bit (i.e., $128 - 3 * 32 = 32$) 
padding element is implicitly inserted into the structure \texttt{Pt} to meet the 128 
bit size requirement for alignment; while on double precision, the hidden 
padding element is 64 bit, and the access to this structure needs two 128 
bit read or write instructions (i.e., $64 * 3 + 64 = 2 * 128$). 

\begin{figure}[htb]
\centering
    \includegraphics[scale = 0.6]{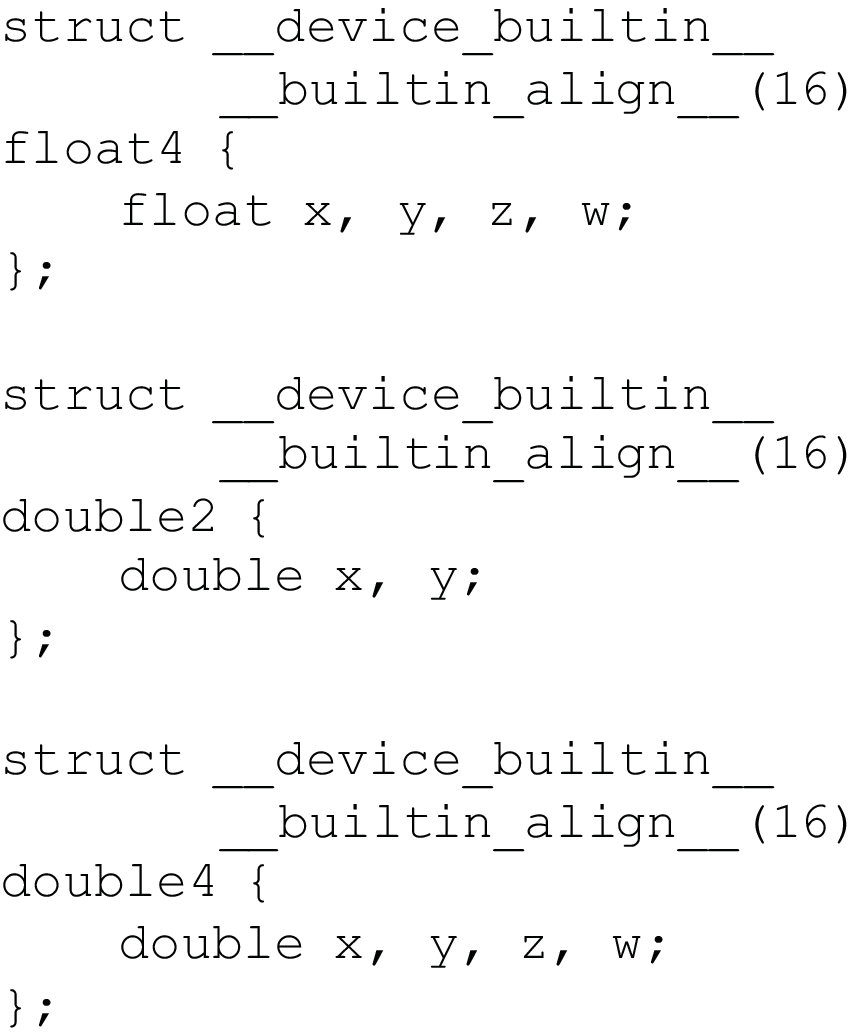}
    \caption{Several build-in data types in CUDA. (The data type \texttt{double4} is aligned into two 16 bytes words)}
    \label{fig:layout:buildin}       
\end{figure}

Another notable issue in exploring the AoaS layout is the use of build-in data types. CUDA has provided various build-in data types; see Figure \ref{fig:layout:buildin} for three examples. The size requirement for alignment is automatically fulfilled for some built-in data types like \texttt{float2}, \texttt{float4}, or \texttt{double2}. 

We also use the build-in types \texttt{float4} and \texttt{double4} to develop a \textit{build-in} group of GPU implementations. This build-in group of GPU implementations is quite easily implemented by replacing the structure \texttt{Pt} with \texttt{float4} or \texttt{double4}. The only difference between the AoaS format data types illustrated in Figure \ref{fig:layout:aoas} and those build-in types shown in Figure \ref{fig:layout:buildin} is that: in the AoaS format data types, a hidden padding element is implicitly added, while the component \texttt{w} is explicitly defined in \texttt{float4} or \texttt{double4} to be used as a padding element. 

We test the build-in group of GPU implementation and compare the performance with that of the AoaS group. We find that there are no remarkable performance gains. Hence, we do not adopt these build-in data types to form combined data types, but choose the user-defined data types to create hybrid types; see Figure \ref{fig:layout:soaos} and Figure \ref{fig:layout:hybrid}.

\subsection{The SoAoS Group of Implementations}
\label{sec:implement:soaos}

When operating on structures residing in global memory, typically there are 
two major optimization strategies \cite{siegel2009}:

\begin{itemize}
\item Accessing consecutive elements to guarantee for coalesced reads.

\item Alignment of data structures to allow for fewer reads.
\end{itemize}

\begin{figure}[htb]
\centering
    \includegraphics[scale = 0.6]{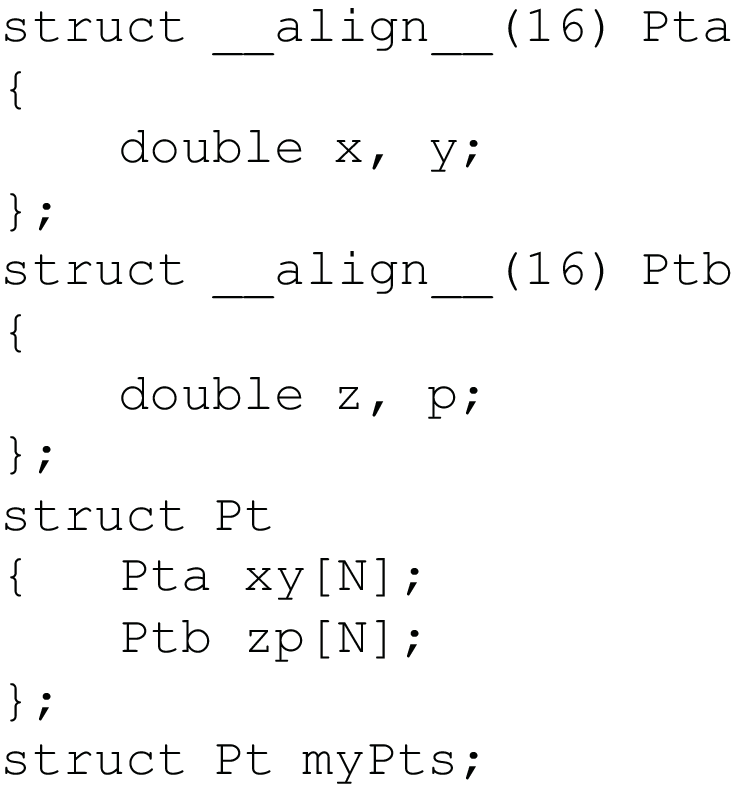}
    \caption{Data layout: Structure of Arrays of aligned Structures (SoAoS)}
    \label{fig:layout:soaos}       
\end{figure}

The above two strategies can generally achieve performance improvements in 
the use of global memory. In order to benefit from both methods, a combined 
data layout, \textit{Structure of Arrays of aligned Structures} (SoAoS), is proposed in \cite{siegel2009}. By organizing aligned 
structures that don't exceed the alignment boundary in multiple arrays, it 
is able to reduce the overall number of issued reads by using 64 or 128 bit 
memory accesses while guaranteeing that all the memory accesses of the 
single threads within the same warp (or half-warp on some devices) are 
coalesced; see Figure \ref{fig:layout:soaos}. Note that the component \texttt{p} in the structure \texttt{Ptb} is just 
an explicit padding element that will never be used in calculating. 

The data structures illustrated in Figure \ref{fig:layout:soaos} are particularly designed for 
double precision. In this paper, we only implement the three GPU 
implementations of the IDW interpolation on double precision according to 
the SoAoS layout and related data structures.

\subsection{The Hybrid Group of Implementations}
\label{sec:implement:hybrid}

\begin{figure}[htb]
\centering
    \includegraphics[scale = 0.6]{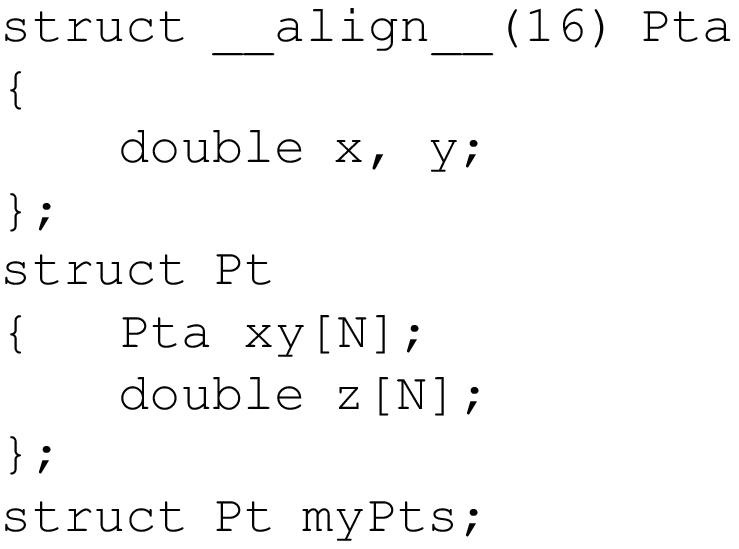}
    \caption{The hybrid data layout by combining AoS and AoV}
    \label{fig:layout:hybrid}       
\end{figure}

The SoAoS layout described above is a combination of the layouts SoA and 
AoS. In this paper, specifically for the IDW interpolation, we also 
introduce another combined data layout which is a combination of the AoS and the 
\textit{Array of Values }(AoV); see Figure \ref{fig:layout:hybrid}. A major difference between this hybrid layout and the 
SoAoS layout is the use of an AoV format array (i.e., \texttt{double z[N]} in Figure \ref{fig:layout:hybrid}) to replace an AoS format array
(i.e, \texttt{Ptb zp[N]} in Figure \ref{fig:layout:soaos}). Another difference is that in this hybrid layout there 
is no explicit or implicit padding element. 

Similar to the layout SoAoS, the hybrid layout is only applicable on double 
precision. Thus, we also implement the three GPU implementations only on 
double precision according to this hybrid layout and related data structures 
illustrated in Figure \ref{fig:layout:hybrid}.

\section{Results and Discussion}
\label{sec:result:and:discuss}

\subsection{Results}
\label{sec:result}

The GPU implementations are evaluated using the NVIDIA GeForce GT640 (GDDR5) 
graphics card and the CUDA 5.5. Note that the GeForce GT640 card with 
memory GDDR5 has the Compute Capability 3.5, while it only has Compute 
Capability 2.1 with the memory DDR3. For each set of the testing data, we 
carry out all GPU implementations on both single precision and double 
precision. 

For the CPU implementations, we directly adopt our previous results that 
were performed on single precision. These results have been presented in \cite{mei2014}; and in this paper, they are directly accepted to be used as the 
baseline. The efficiency performance of all GPU implementations is 
benchmarked by comparing to the baseline results.

As described in \cite{mei2014}, for each GPU implementation, we tested two 
different forms that have different values of the power parameter $p$. In the 
first form, the power $p$, see Equation 1, is set to an integer value 2, while 
this value is set to 3.0 in the second form. In this paper, we only consider 
the first form (i.e., $p = 2$).

The input of the IDW interpolation is the coordinates of data points and 
prediction points. The performance of the CPU and GPU implementations may 
differ due to different sizes of input data \cite{hanzer2012,hennebohl2011}. However, the 
motivation of this work is focused on evaluating the performance impact of 
different data layouts; thus, we only consider a special situation where the 
numbers of prediction points and data points are identical.

We create five groups of sizes, i.e., 10K, 50K, 100K, 500K, and 1000K 
(1K=1024). And five tests are performed by setting the numbers of both the 
data points and prediction points as the above listed five groups of sizes.

\subsubsection{Single Precision}
\label{sec:result:single}

\begin{figure}[htb]
\centering
\subfigure[]{
\label{fig:single:a}       
\includegraphics[scale = 0.35]{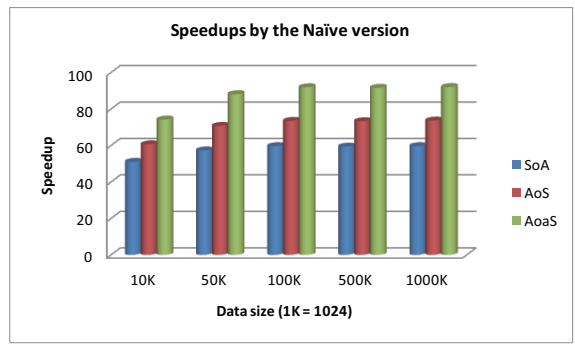}
}
\hspace{1em}
\subfigure[]{
\label{fig:single:b}       
\includegraphics[scale = 0.35]{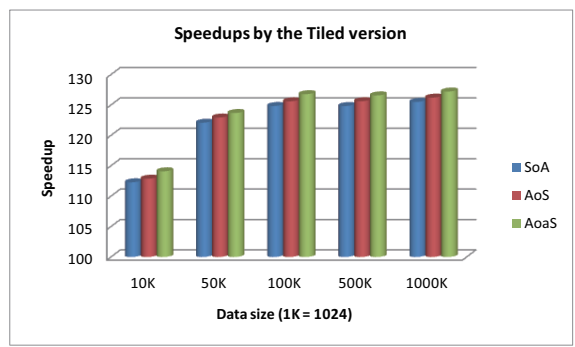}
}
\hspace{1em}
\subfigure[]{
\label{fig:single:c}       
\includegraphics[scale = 0.35]{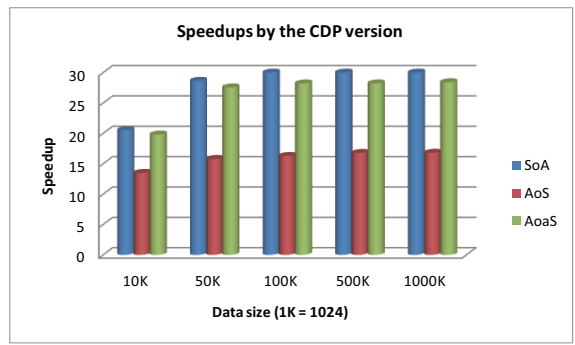}
}
\caption{Performance of GPU implementations on single precision}
\label{fig:single}       
\end{figure}

On single precision, we implement those three GPU implementations of the IDW 
interpolation using three types of data layouts, including the SoA, the AoS, 
and the AoaS. The benchmark results (i.e., speedups generated 
by comparing to the baseline CPU results) of the na\"{\i}ve version, the 
tiled version, and the CDP version are shown in Figure \ref{fig:single}.

According to the results generated in above three experimental tests, we have found 
that: for both the na\"{\i}ve and tiled implementations, the layout AoaS 
achieves the best performance and the layout SoA obtains the worst results; 
see Figures \ref{fig:single:a} and \ref{fig:single:b}. However, for the CDP version, the layout SoA gets 
the best performance, and the second best is the layout AoaS, while the AoS 
layout leads the worst results; see Figure \ref{fig:single:c}.

\subsubsection{Double Precision}
\label{sec:result:double}

\begin{figure}[htb]
\centering
\subfigure[]{
\label{fig:double:a}       
\includegraphics[scale = 0.35]{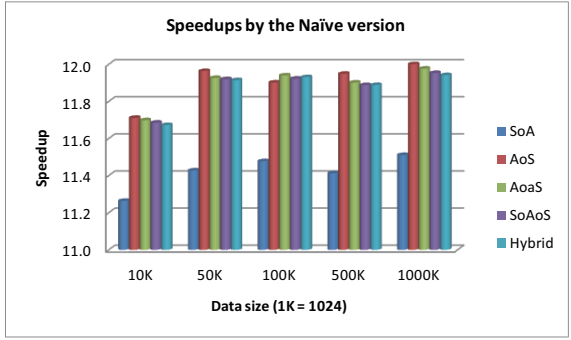}
}
\hspace{1em}
\subfigure[]{
\label{fig:double:b}       
\includegraphics[scale = 0.35]{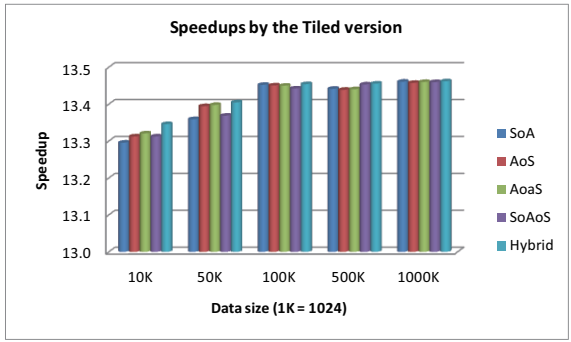}
}
\hspace{1em}
\subfigure[]{
\label{fig:double:c}       
\includegraphics[scale = 0.35]{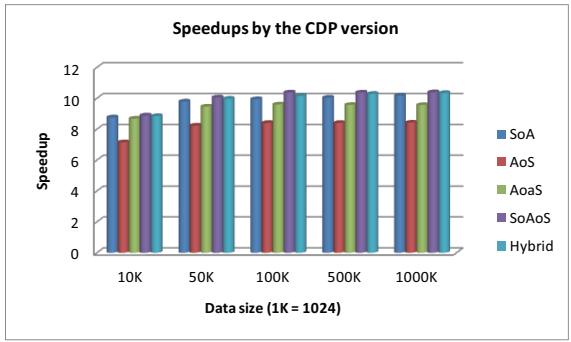}
}
\caption{Performance of GPU implementations on double precision}
\label{fig:double}       
\end{figure}

On double precision, we implement two groups of those three GPU implementations using 
additional two types of combined data layouts, the SoAoS and the Hybrid; we 
also implement the GPU implementations using the layouts SoA, AoS, and AoaS. 
The experimental results in this case are presented in Figure \ref{fig:double}.

For the na\"{\i}ve version, the speedups generated by the GPU implementation 
according to the layout SoA are the lowest, while the other four layouts 
achieve almost the same performance although the speedups are slightly 
varied; see Figure \ref{fig:double:a}.

For the tiled version, all of the five different data layouts obtain nearly the 
same performance. There are only several slight differences among those speedups 
; see Figure \ref{fig:double:b}.

For the CDP version, the layout AoS leads the worst performance; and the 
second worst results are generated by the layout AoaS. The other three 
layouts including the SoA, the SoAoS, and the Hybrid obtain almost the same 
performance; see Figure \ref{fig:double:c}.

\subsection{Discussion}
\label{sec:discussion}

Recently, the GPU-computing programming models such as CUDA are popularly 
used to speed up various scientific applications. However, fully utilizing 
the specific features of the underlying GPU architecture is still a 
challenging work. One of the most important responsibilities of a programmer 
is to maximize the efficiency performance by optimizing the memory hierarchy 
in GPU-computing. 

The data layout in memory is a critical issue in developing efficient GPU 
code. Several efforts have been carried out to analyze \cite{giles2013,siegel2009,strzodka2012} or 
transform \cite{mistry2011,strzodka2011,sung2012} different types of data layouts.

Based upon our previous work \cite{mei2014}, in this paper we focus on evaluating 
the performance impact of different data layouts on the IDW interpolation. 
First, we develop several sets of the GPU implementations of the standard 
IDW interpolation according to a set of data layouts, and then test their 
efficiency performance on both single precision and double precision. 

On single precision, we implement three groups of the GPU implementations using three 
types of data layouts, including the SoA, the AoS, and the AoaS. We find that: 
for the AoS and AoaS layouts, the second one always obtains better 
performance than the first. This positive impact on performance is due to 
minimizing the number of memory transactions by aligning the data 
structures. 

We also observe that: for all the three versions of GPU implementations, the 
performance impact due to the use of the AoaS over the AoS for both the 
na\"{\i}ve version and the CDP version is much more significant than that 
for the tiled version; see Figure \ref{fig:single}.

The above performance result is perhaps because of the effective 
optimization in the use of global memory by minimizing the number of memory 
transactions. In both the na\"{\i}ve version and the CDP version, each 
thread needs to read the coordinates of all data points; in other words, the 
coordinates of all data points are needed to be read \textit{n} times, where \textit{n} is the 
number of predication points. In contrast, the coordinates of all data 
points are only needed to be read (\textit{n} / threadsPerBlock) times due to 
accepting the optimization strategy ``\textit{tiling}''. Thus, there are much 
more global memory accesses in both the na\"{\i}ve and the CDP versions than 
that in the tiled version. And the impact of optimizing the use of global 
memory by minimizing the number of transactions on a larger number of global 
memory accesses is obviously more significant than that on a smaller number 
of global memory accesses. 

The above two layouts (AoS and AoaS) achieve higher speedups than the layout 
SoA for both the na\"{\i}ve and tiled implementations, but get lower 
speedups than the SoA for the CDP implementation. We cannot explain this 
strange behavior. Perhaps this behavior is due to the nested parallelism 
when programming with CUDA dynamic parallelism.

Another notable issue in exploring the AoaS layout is the use of build-in data types. CUDA has provided various build-in data types. The size requirement for alignment is automatically fulfilled. 
Compared to those user-defined data types in AoaS formant (see Figure \ref{fig:layout:aoas}), we have found the counterparts of the build-in data types provided by CUDA do not achieve notable advantages. In addition, the user-defined AoaS data types are suggested to be used for the convenience in programming.

On double precision, we also observe some performance results that are 
as the same as those on single precision:

\begin{enumerate}
\item For the na\"{\i}ve version, both the layouts AoS and AoaS are better than the SoA.

As explained above, this positive result is because of aligning the data 
structures to allow for fewer reads or writes. 

\item For the tiled version, those three layouts, SoA, AoS, and AoaS achieve almost the same performance. 

This result is due to the fact that the accesses to global memory have been 
optimized using the strategy ``tiling'' and the impact of different data 
layouts on accessing global memory is not significant.

\item For the CDP version, the layout SoA still obtains best results when compared to the layouts AoS and AoaS.

We cannot give reasonable explanations for this strange behavior. We guess 
that the coalesced access to global memory in nested parallelism (CUDA 
dynamic parallelism) has a very positive performance impact. 

\end{enumerate}

Furthermore, we find several additional results on double precision. 

\begin{enumerate}
\item For the na\"{\i}ve version, all the data layouts except the SoA achieve nearly the same speedups. Noticeably, among these four layouts, i.e., the AoS, the AoaS, the SoAoS, and the Hybrid, the best one is the AoS, in which the alignment is not used. 

This result is perhaps due to two reasons: the first is that the aligning for data structures on double precision is not as effective as that on single precision (see Figure \ref{fig:single:a} and Figure \ref{fig:double:a}); the second potential cause is that there is probably a performance penalty when aligning data structures on double precision. However, the advantage of the AoS layout is not obvious.

\item For all the three versions, the performance differences between the SoAoS layout and the Hybrid layout are quite small. This illustrates that the use of the AoS or the AoV in a combined layout on double precision does not lead to heavy impact on performance. 
\end{enumerate}

Considering the overall performance on single and double precision, we 
recommend that: for both the na\"{\i}ve version and the tiled version, the 
best choice is the data layout AoaS, while the layout SoA is the best one 
for the CDP version. From the perspective of GPU performance in practical applications, the layout 
AoaS is suggested to be the only option since that the tiled version is the 
fastest one among the three versions of GPU implementations. 

In this paper, all the experimental tests are performed and evaluated on a 
single GPU. In some related work \cite{guan2010,huang2011}, efficient implementations of 
the IDW interpolation were developed on the platforms of multiple GPUs or on
clusters. When intending to benefit from multi-GPUs or clusters, it is 
needed to carefully analyze and select the optimal data layout. Future work 
should therefore include the implementation of the IDW interpolation and the 
performance evaluation of different data layouts under the environment of 
multi-GPUs or clusters.

\section{Conclusion}
\label{sec:conclusion}

We have redesigned and improved the CDP version of the GPU implementations 
of the standard IDW interpolation algorithm by exploiting the feature CUDA 
Dynamic Parallelism. We have demonstrated that the improved CDP version has 
the speedups of 2.9x and 1.5x over the original CDP version when the power 
parameter $p$ is set to 2 and 3.0, respectively. In further, in order to 
evaluate the performance impact of different data layouts, we have 
implemented the na\"{\i}ve version, the tiled version, and the improved CDP 
version based upon three basic layouts (SoA, AoS, and AoaS) and two combined 
layouts. We have observed that: (1) For both the na\"{\i}ve version and 
tiled version, the layouts AoS and AoaS achieve better performance than the 
layout SoA; (2) For the improved CDP version, the layout SoA is the best 
choice among the three basic layouts; (3) For the two combined data layouts, 
there are no notable performance gains when compared to those three basic 
layouts. We recommend that: in practical applications, the layout AoaS is 
the best choice since the tiled version is the fastest one among the three 
versions of GPU implementations, especially on single precision.

\begin{flushleft}
\textbf{Conflict of Interests}
The authors declare that there is no 
conflict of interests regarding the publication of this article.
\end{flushleft}

\begin{acknowledgements}
The authors are grateful to the anonymous referee 
for helpful comments that improve this paper.
\end{acknowledgements}

\bibliographystyle{ieeetrans} 
\bibliography{myBib}   

\end{document}